\documentclass[]{article}
\usepackage{graphics,epsfig}
\title{Quantum Structures: An Attempt to Explain the Origin of their Appearance in Nature\footnote{Published as:
Aerts, D., 1995, ``Quantum structures: an attempt to explain the origin of their appearance in nature", {\it
International Journal of Theoretical Physics}, {\bf 34}, 1165.}}
\author{Diederik Aerts}
\date{}
\font\smallroman=cmr10 at 8pt

\newtheorem{theorem}{Theorem}
\newtheorem{definition}{Definition}

\begin{document}
\maketitle
\centerline{FUND and CLEA,}
\centerline {Brussels Free University, Krijgskundestraat 33,}
\centerline {1160 Brussels, Belgium,}
\centerline {e-mail: diraerts@vub.ac.be}
\begin{abstract}
\noindent We explain
the quantum structure as due to the presence of two effects, (a) a 
real change of state
of the entity under influence of the measurement and, (b) a lack 
of knowledge about a deeper
deterministic reality of the measurement process. We present a 
quantum machine, where we can
illustrate in a simple way how the quantum structure arises as a 
consequence of the two
mentioned effects. We introduce a parameter $\epsilon$ that 
measures the size of the lack of
knowledge on the measurement process, and by varying this
parameter, we describe a continuous evolution from a quantum 
structure (maximal lack of
knowledge) to a classical structure (zero lack of knowledge). We 
show that for intermediate
values of $\epsilon$ we find a new type of structure, that is 
neither quantum
nor classical. We apply the model that we have introduced to 
situations of lack of knowledge
about the measurement process appearing in other regions of 
reality. More specifically we
investigate the quantum-like structures that appear in the 
situation of psychological
decision processes, where the subject is influenced during the 
testing, and forms some of
his opinions during the testing process. Our conclusion is that in 
the light of this
explanation, the quantum  probabilities are epistemic and not 
ontological, which means
that quantum mechanics is compatible with a determinism of the 
whole.
\end{abstract}

\section{A macroscopic machine producing quantum structure}

Before we come to identify the origin of the appearance 
of quantum structures in
nature, we want to expose a macroscopic machine that produces 
quantum structure. We shall
make use intensively of the internal functioning of this machine 
to demonstrate our general
explanation.

The machine that we consider consists of a physical entity 
$S$ that is a
point particle $P$ that can move on the surface of a sphere, 
denoted $surf$, with center $O$
and radius $1$. The unit-vector $v$ where the particle is located 
on $surf$
represents the state $p_v$ of the particle (see Fig. 1,a). For 
each point $u \in surf$, we
introduce the following experiment $e_u$. We consider the 
diametrically opposite point $-u$,
and install a piece of elastic of length 2, such that it is fixed 
with one of its end-points
in $u$ and the other end-point in $-u$. Once the elastic is 
installed, the particle $P$ falls
from its original place $v$ orthogonally onto the elastic, and 
sticks on it (Fig 1,b).

\vskip 0.7 cm
\hskip 2.2 cm \includegraphics{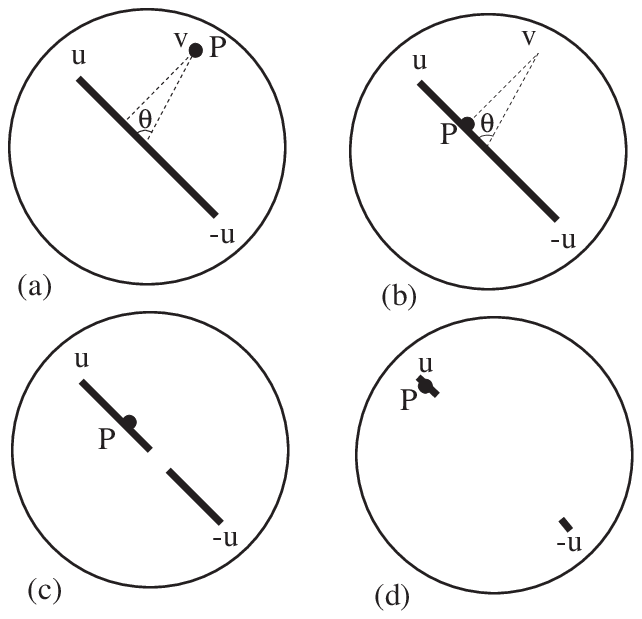}

\begin{quotation}
\noindent \baselineskip= 9 pt \smallroman Fig. 1 : A representation of the 
quantum machine. In (a) the
physical entity $\scriptstyle P$ is in state $\scriptstyle p_v$ in 
the point $\scriptstyle
v$, and the elastic corresponding to the experiment $\scriptstyle 
e_{u}$ is installed
between the two diametrically opposed points $\scriptstyle u$ and 
$\scriptstyle -u$. In (b)
the particle $\scriptstyle P$ falls orthogonally onto the elastic 
and sticks to it. In (c)
the elastic breaks and the particle $\scriptstyle P$ is pulled 
towards the point
$\scriptstyle u$, such that (d) it arrives at the point 
$\scriptstyle u$, and the
experiment $\scriptstyle e_u$ gets the outcome $\scriptstyle 
o^u_1$.
\end{quotation}
Then
the elastic breaks and the particle $P$, attached to one of the 
two pieces of the
elastic (Fig. 1,c), moves to one of the two end-points $u$ or $-u$ 
(Fig. 1,d). Depending on whether the particle $P$ arrives in $u$ (as in 
Fig. 1) or in $-u$,
we give the outcome $o^u_1$ or $o^u_2$ to $e_u$. In Figure 2 we 
represent the experimental
process connected to $e_u$ in the plane where it takes place, and 
we can easily
calculate the probabilities corresponding to the two possible 
outcomes. In order to do so we
remark that the particle $P$ arrives in $u$ when the elastic 
breaks in a point of the interval
$L_1$, and arrives in $-u$ when it breaks in a point of the 
interval $L_2$ (see Fig. 2). We
make the hypothesis that the elastic breaks uniformly, which means 
that the probability
that the particle, being in state $p_v$, arrives in $u$, is given 
by the length of $L_1$
(which is $1+ cos\theta$) divided by the length of the total 
elastic (which is 2). The
probability that the particle in state $p_v$ arrives in $-u$ is 
the length of $L_2$ (which is
$1-cos\theta$) divided by the length of the total elastic. If we 
denote these probabilities
respectively by $P(o^u_1, p_v)$ and $P( o^u_2, p_v)$ we have:

\begin{eqnarray}
P(o^u_1, p_v) &=& {{1+cos\theta}\over 2} = cos^2{\theta\over 2} \\
P(o^u_2, p_v) &=& {{1-cos\theta}\over 2} = sin^2{\theta\over 2}
\end{eqnarray}
The probabilities that we find in this way are 
exactly the quantum probabilities
for the spin measurement of a spin 1/2 quantum entity, which means 
that we can describe this
macroscopic machine by the ordinary quantum formalism with a two 
dimensional
complex Hilbert space as the carrier for the set of states of the 
entity.

\vskip 0.7 cm
\hskip 3.8 cm \includegraphics{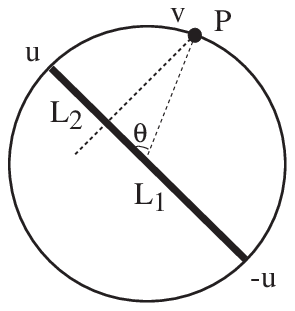}

\begin{quotation}
\noindent \baselineskip= 9 pt \smallroman Fig. 2 : A representation of the 
experimental process in the
plane where it takes place. The elastic of length 2, corresponding 
to the experiment
$\scriptstyle e_u$, is installed between $\scriptstyle u$ and 
$\scriptstyle -u$. The
probability, $\scriptstyle P(o^u_1, p_v)$, that the particle 
$\scriptstyle P$ ends up
in point $\scriptstyle u$ is given by the length of the piece of 
elastic $\scriptstyle L_1$
divided by the length of the total elastic. The probability, 
$\scriptstyle P(o_2^u, p_v)$,
that the particle $\scriptstyle P$ ends up in point $\scriptstyle 
-u$ is given by the length
of the piece of elastic $\scriptstyle L_2$ divided by the length 
of the total elastic.
\end{quotation}

\section{Quantum Structures}
Already from the advent of quantum mechanics it was 
known that the structure of
quantum theory is very different from the structure of the earlier 
existing classical
theories. This structural difference has been expressed and 
studied in different mathematical
categories, and we mention here some of the most important ones : 
(1) if one considers the
collection of properties (experimental propositions) of a physical 
entity, then it has the
structure of a Boolean lattice for the case of a classical entity, 
while it is non-Boolean
for the case of a quantum entity (Birkhoff and Von Neumann 1936, 
Jauch 1968, Piron 1976), (2)
for the probability model, it can be shown that for
a classical entity it is Kolmogorovian, while for a quantum entity 
it is not
(Foulis and Randall 1972, Randall and Foulis 1979, 1983 , Gudder 
1988, Accardi 1982, Pitovski
1989), (3) if the collection of observables is considered, a 
classical entity gives rise to a
commutative algebra, while a quantum entity doesn't (Segal 1947 , 
Emch 1984).

The presence of these deep structural differences between 
classical theories and quantum
theory has contributed strongly to the already earlier existing 
belief that classical
theories describe the ordinary 'understandable' part of reality, 
while quantum theory
confronts us with a part of reality (the micro-world) that is 
impossible to understand.
Therefore still now there is the strong paradigm that {\it quantum 
mechanics cannot be
understood}. The example of our macroscopic machine with a quantum 
structure challenges
this paradigm, because obviously the functioning of this machine 
can be understood. The aim
of this paper is to show that the main part of the quantum 
structures can indeed be explained
in this way and the reason why they appear in nature can be 
identified. In this paper we
shall analyze this explanation, that we have named the 'hidden 
measurement approach', in the
category of the probability models. We refer to (Aerts and Van 
Bogaert 1992, Aerts, Durt and
Van Bogaert 1993, Aerts, Durt, Grib, Van Bogaert and Zapatrin 
1993, Aerts 1994, Aerts and Durt
1994a) for an analysis of this explanation in other categories.

The original development of probability theory was aiming at a 
formalization of the
description of a probability that appears as the consequence of 
{\it a lack of
knowledge}. The probability structure appearing in situations of 
lack of knowledge was
axiomatized by Kolmogorov and such a probability model is now 
called Kolmogorovian. Since
the quantum probability model is not Kolmogorovian, it has now 
generally
been accepted that the quantum probabilities are {\it not} a 
description of a {\it lack of
knowledge}. Sometimes this conclusion is formulated by stating 
that the quantum probabilities
are {\it ontological} probabilities, as if they would be present 
in reality itself. In the
hidden measurement approach we show that the quantum probabilities 
can be explained as being
due to a {\it lack of knowledge}, and we prove that what 
distinguishes quantum
probabilities from classical Kolmogorovian probabilities is the 
{\it nature of this lack of
knowledge}. Let us go back to the quantum machine to illustrate 
what we mean.

If we consider again our quantum machine (Fig. 1 and Fig. 2), and 
look for the
origin of the probabilities as they appear in this example, we can 
remark that the
probability is entirely due to a {\it lack of knowledge} about the 
measurement process.
Namely the lack of knowledge of where exactly the elastic breaks 
during a measurement.
More specifically, we can identify two main aspects of the 
experiment $e_u$ as it appears
in the quantum machine. 

\begin{enumerate}
\item The experiment $e_u$ effects a real change on the 
state $p_v$
of the entity $S$. Indeed, the state $p_v$ changes into one of the 
states $p_u$ or $p_{-u}$
by the experiment $e_u$.

\item The probabilities appearing are due to a {\it 
lack of knowledge} about a
deeper reality of the individual measurement process itself, 
namely where the elastic breaks.
\end{enumerate}

\noindent These two effects give rise to quantum-like
structures. The lack of knowledge about a deeper reality of the 
individual measurement
process we have referred to as the presence of 'hidden 
measurements' that operate
deterministically in this deeper reality (Aerts 1986, 1987, 1991), 
and that is the origin of
the name that we gave to this approach. A consequence of this 
explanation is that
quantum structures shall turn out to be present in many other 
regions of reality where the
two mentioned effects appear. We think of the many situations in 
the human sciences,
where generally the measurement disturbs profoundly the entity 
under study, and where almost
always a lack of knowledge about the deeper reality of what is 
going on during this
measurement process exists. In the final part of this paper we 
give some examples of quantum
structures appearing in such situations.

\section{Quantum, Classical and Intermediate}

If the quantum structure can be explained by the 
presence of a lack of knowledge on
the measurement process, we can go a step further, and wonder what 
types of structure arise
when we consider the original models, with a lack of knowledge on 
the measurement process, and
introduce a variation of the magnitude of this lack of knowledge. 
We have studied the
quantum machine under varying 'lack of knowledge', parameterizing 
this variation by a number
$\epsilon \in [0,1]$, such that $\epsilon = 1$ corresponds to the 
situation of maximal lack
of knowledge, giving rise to a quantum structure, and $\epsilon = 
0$ corresponds to the
situation of zero lack of knowledge, generating a classical 
structure, and other values of
$\epsilon$ correspond to intermediate situations, giving rise to a 
structure that is neither
quantum nor classical (Aerts, Durt and Van Bogaert 1992, 1993). It 
is this model that we have
called the $\epsilon$-model, and we want to introduce it again in 
this paper.

We start from the quantum machine, but introduce now different 
types of elastic.
An $\epsilon, d$-elastic consists of three different parts: one 
lower part where it is
unbreakable, a middle part where it breaks uniformly, and an upper 
part where it is again
unbreakable. By means of the two parameters $\epsilon \in [0,1]$ 
and $d \in [-1+\epsilon,
1-\epsilon]$, we fix the sizes of the three parts in the following 
way. Suppose
that we have installed the $\epsilon, d$-elastic between the 
points $-u$ and $u$ of the
sphere. Then the elastic is unbreakable in the lower part from $-
u$ to $(d-\epsilon) \cdot
u$, it breaks uniformly in the part from $(d-\epsilon) \cdot u$ to 
$(d+\epsilon) \cdot u$,
and it is again unbreakable in the upper part from $(d+\epsilon) 
\cdot u$ to $u$ (see Fig. 3).

\vskip 0.7 cm
\hskip 3.8 cm \includegraphics{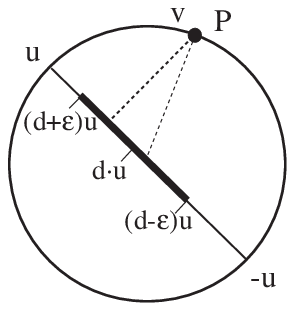}

\begin{quotation}
\noindent \baselineskip= 9 pt \smallroman Fig. 3 :  A representation of the 
experiment $\scriptstyle
e^\epsilon_{u, d}$. The elastic breaks uniformly between the 
points
$\scriptstyle (d-\epsilon) u$ and $\scriptstyle (d +\epsilon) u$, 
and is unbreakable
in other points.
\end{quotation}
 An $e_u$ experiment performed by means of an $\epsilon, 
d$-elastic shall be denoted
by $e^\epsilon_{u,d}$. We have the following cases:
\smallskip
\noindent {\it (1)} $v \cdot u \le d-\epsilon$. The particle 
sticks to the lower part of
the $\epsilon, d$-elastic, and any breaking of the elastic pulls 
it down to
the point $-u$. We have $P^\epsilon(o^u_1, p_v) = 0$ and 
$P^\epsilon(o^u_2, p_v) = 1$.

\smallskip
\noindent {\it (2)} $d-\epsilon < v \cdot u < d+\epsilon$. The 
particle falls onto the
breakable part of the $\epsilon, d$-elastic. We can easily 
calculate the transition
probabilities and find:
\begin{eqnarray}
P^\epsilon(o_1^u, p_v) &=& {1 \over 2\epsilon}( v \cdot u -d 
+\epsilon) \label{equation3} \\
P^\epsilon(o_2^u, p_v) &=& {1 \over 2\epsilon}(d+\epsilon - v
\cdot u) \label{equation4}
\end{eqnarray}
{\it (3)} $d+\epsilon \le v \cdot u$. The particle
falls onto the upper part of the $\epsilon,d$-elastic, and any 
breaking of the elastic pulls
it upwards, such that it arrives in $u$. We have 
$P^\epsilon(o_1^u, p_v) = 1$ and
$P^\epsilon(o_2^u,  p_v) = 0$.

\section{Probabilities appearing in physical situations}

If we want to analyze the structure of the quantum 
probability model in the light
of axioms that have been formulated for classical probability 
theory, we first have to be
very specific about the situation that we consider. In physics 
(and hence also in
quantum mechanics and classical mechanics) we consider a situation 
where we have a
physical entity $S$ that can be in different states $p, q, r, 
...$, and we'll denote the set
of states by $\Sigma$. On this physical entity $S$, in a certain 
state $p$, we perform
experiments $e, f, g, ...$, that respectively have sets of 
possible outcomes $O_e$, $O_f$,
$O_g$, ....Let us denote the collection of all relevant 
experiments by ${\cal E}$. There
are different places in which probability appears in this scheme.

\subsection{The probability connected to the states}
In many occasions it is not
possible to prepare the entity $S$ in such a way that we know in 
which state it is before we
start an experiment. We can only prepare it such that we are left 
with a situation of 'lack of
knowledge' about the state of the entity. This situation of lack 
of knowledge is described
by means of a probability measure $\mu : {\cal B}(\Sigma) \to 
[0,1]$ on the set of states,
such that ${\cal B}(\Sigma)$ is a $\sigma$-algebra of measurable 
subsets of $\Sigma$, and for
$K \in {\cal B}(\Sigma)$ we have that $\mu(K)$ is the probability 
that the state of the entity
$S$ is in the subset $K$. We have (a)
$\mu(\Sigma) = 1$, and (b) $\mu(\cup_iK_i) = \sum_i\mu(K_i)$ for 
sets $K_i$ such that $K_n
\cap K_m = \emptyset$ for $n \not= m$. What we call 'states' $p, 
q, r, ...$ are often called
'pure states', and what we call 'situations of lack of knowledge 
on the states' $\mu, \nu,
...$ are often called 'mixed states'.

\subsection{The probability connected to the experiments}
Even when the entity $S$ is in a state $p$, and an experiment $e$ 
is performed, probability,
defined as the limit of the relative frequency connected to an 
outcome $o_k \in O_e$, appears.
For a fixed state $p$, the probability that an experiment $e$ 
gives an outcome in a subset
$A^e \subset O_e$, denoted by $P(A^e, p)$, can be described as a 
probability
measure on the outcome set $O_e$ of the experiment $e$. Hence a 
map $P : {\cal B}(O_e)
\times \Sigma \to [0,1]$ such that ${\cal B}(O_e)$ is a collection 
of measurable subsets of
$O_e$ and (a) $P(O_e, p) = 1$, and (b) $P(\cup A_k^e, p) = 
\sum_kP(A_k^e, p)$ for sets
$A_k^e$ such that $A_i^e \cap A_j^e = \emptyset$ for $i \not= j$.

\subsection{The general probability}

Most of the time we measure a probability in the 
laboratory that contains both
just mentioned probabilities. It is the probability that in a 
situation $\mu$ of lack of
knowledge on the states, an experiment $e$ gives an outcome in a 
subset $A^e \subset O_e$, and
we'll denote it by $P(A^e, \mu)$. This probability is, for a given 
experiment $e$, a map $P:
{\cal B}(O_e) \times {\cal M}(\Sigma) \to [0,1]$, where ${\cal 
M}(\Sigma)$ is the set of
probability measures on $\Sigma$.

\subsection{The eigenstate sets and the possibility-state 
sets}
As we have done in (Aerts 1994) we introduce for an 
experiment $e$ the eigenstate
sets, as maps $eig : {\cal P}(O_e) \to {\cal P}(\Sigma)$, where 
for $A^e \subset O_e$ we
have :
\begin{equation}
eig(A^e) = \{p\ \vert\ p \in \Sigma, {\rm if\ } S {\rm\ is\ in\ 
} p {\rm\ the\
outcome\ of}\ e {\rm\ occurs\ certainly\ in\ } A^e\} 
\end{equation}
\noindent We also introduce the
possibility-state sets, as maps $pos : {\cal P}(O_e) \to {\cal 
P}(\Sigma)$, where for $A^e
\subset O_e$ we have : 
\begin{equation}
pos(A^e) = \{p\ \vert\ p \in \Sigma, {\rm if\ } S {\rm\ is\ in\ 
} p
{\rm the\ outcome\ of}\ e {\rm\ occurs\ possibly\ in\ } 
A^e\}
\end{equation}
Clearly we always have :
\begin{equation}
eig(A^e) \subset pos(A^e)
\end{equation}
\begin{theorem} \label{theorem1} We consider an entity $S$ in a 
situation with lack
of knowledge about the states described by the probability measure 
$\mu$ on the state
space. For an arbitrary experiment $e$ and set of outcomes $A^e 
\subset O_e$ we have:
\begin{equation}
\mu(eig(A^e)) \le P(A^e, \mu) \le \mu(pos(A^e)) \label{equation1}
\end{equation}
\end{theorem} 
{\sl Proof:} As defined we have that $\mu(eig(A^e))$ is the 
probability that the state of the
entity is in the subset $eig(A^e)$. If the state is in $eig(A^e)$, 
the experiment $e$ gives
with certainty an outcome in $A^e$, and therefore $\mu(eig(A^e)) 
\le P(A^e, \mu)$. As defined
we have that $\mu(pos(A^e))$ is the probability that the state of 
the entity is in $pos(A^e)$.
If the state is in $pos(A^e)$ the experiment $e$ has a possible 
outcome in $A^e$, and
therefore $P(A^e, \mu) \le \mu(pos(A^e))$. 

\medskip
\noindent We shall show in the next section that for a classical
probability model, the two inequalities become equalities. But 
first we want to illustrate
all these concepts on the quantum machine.

\section{Illustration on the quantum machine}

It is easy to see how these concepts are defined for the 
quantum
machine (and also for the $\epsilon$-model). For a considered 
experiment $e_u$ (or
$e^\epsilon_{u,d}$ in the $\epsilon$-model), we have an outcome 
set $O_u = \{o^u_1,
o^u_2\}$. The set of states is $\Sigma = \{p_v\ \vert\ v \in 
surf\}$. The situations with
lack of knowledge about the states are described by probability 
measures on the surface of
the sphere. We have also described $P^\epsilon(A_u, p_v)$ for an 
arbitrary $A_u \subset O_u$
(see (\ref{equation3}) and (\ref{equation4})). For the eigenstate sets and the possibility-
state sets we have (see Fig. 4.) :
\begin{eqnarray}
eig^\epsilon(\{o^u_1\}) &=& \{p_v \ \vert\ d+\epsilon \le v \cdot 
u \} \\
eig^\epsilon(\{o^u_2\}) &=& \{p_v \ \vert\ v \cdot u \le d-\epsilon 
\} \\
pos^\epsilon(\{o^u_1\}) &=& \{p_v \ \vert\ d-\epsilon < v \cdot u 
\} \\
pos^\epsilon(\{o^u_2\}) &=& \{p_v \ \vert\ v \cdot u < d+\epsilon \} 
\end{eqnarray}

\vskip 0.7 cm
\hskip 2.2 cm \includegraphics{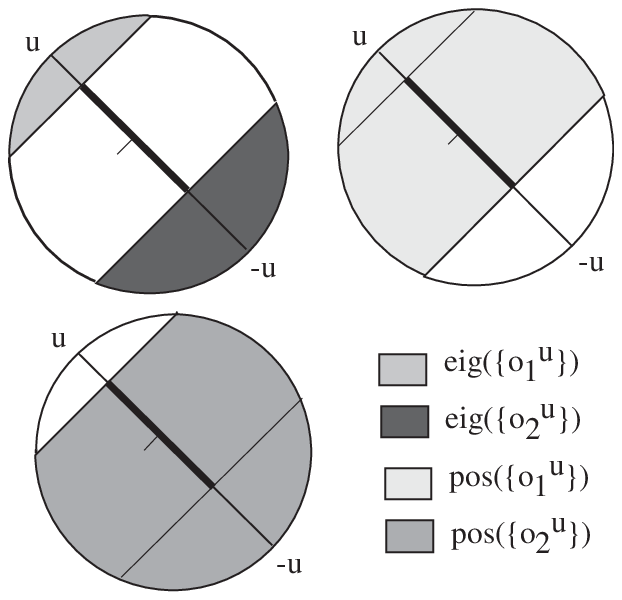}

\begin{quotation}
\noindent \baselineskip= 9 pt \smallroman Fig. 4 : We have represented the 
eigenstate sets
$\scriptstyle eig^\epsilon(\{o_1^u\})$ and $\scriptstyle 
eig^\epsilon(\{o_2^u\})$. If the state of the entity
(the position of the particle $\scriptstyle P$) is in 
$\scriptstyle eig^\epsilon(\{o_1^u\})$
[or in $\scriptstyle eig^\epsilon(\{o_2^u\})$], then the 
experiment $e^\epsilon_{u,d}$ gives with
certainty the outcome $o^u_1$ [or with certainty the outcome 
$o^u_2$]. We also have
represented the possibility state sets, $\scriptstyle 
pos^\epsilon(\{o_1^u\})$ [$\scriptstyle
pos^\epsilon(\{o_2^u\})$], which is the collection of states where 
the entity gives a possible outcome
$o^u_1$ [$o^u_2$].
\end{quotation}
We can consider the following specific cases :

\subsection{The quantum situation ($\epsilon = 1$)}
For $\epsilon$ = 1 we always have $d$ = 0, and the 
$\epsilon$-model reduces to the
original quantum machine that we introduced in section 1. It is a 
model for the spin of a
spin 1/2 quantum entity. The transition probabilities are the same 
as the ones related to the
outcomes of a Stern-Gerlach spin experiment on a spin 1/2 quantum 
particle, of which the
quantum-spin-state in direction $v = (cos\phi sin\theta,$ $ 
sin\phi sin\theta, cos\theta)$,
denoted by $\bar {\psi_v}$, and the experiment $e_u$ corresponding 
to the spin experiment in
direction $u = (cos\beta sin\alpha,$ $sin\beta sin\alpha, 
cos\alpha)$, is described
respectively by the vector and the self adjoint operator 

\begin{eqnarray}
\psi_v 
&=& (e^{-i\phi/2}cos\theta/2,
e^{i\phi/2}sin\theta/2) \\
H_u &=& {1\over 2} 
\pmatrix{cos\alpha &
e^{-i\beta}sin\alpha \cr e^{i\beta}sin\alpha & -cos\alpha \cr} 
\end{eqnarray}
of a
two-dimensional complex Hilbert space. For the eigenstate sets and 
possibility-state sets we
find: 
\begin{eqnarray}
eig^\epsilon(\{o^u_1\}) &=& \{p_u\} \\
eig^\epsilon(\{o^u_2\}) &=& \{p_{-u}\} \\
pos^\epsilon(\{o^u_1\}) &=& \{p_v\ \vert\ v\not= -u\} \\
pos^\epsilon(\{o^u_2\}) &=&
\{p_v\ \vert\ v\not= u\}
\end{eqnarray}
Suppose that we 
consider a situation with lack
of knowledge about the state, described by a uniform probability 
distribution $\mu$ on the
sphere, which corresponds to a random distribution of the point on 
the sphere. Then we can
easily calculate the following probabilities :
\begin{eqnarray}
P^\epsilon(\{o^u_1\}, \mu) &=& {1\over 2} \\
P^\epsilon(\{o^u_2\}, \mu)
&=& {1\over 2}
\end{eqnarray}
On the other hand, we have : 
\begin{eqnarray}
\mu(eig^\epsilon(\{o^u_1\})) &=& 0 \\
\mu(eig^\epsilon(\{o^u_2\})) &=& 0 \\
\mu(pos^\epsilon(\{o^u_1\})) &=& 1 \\
\mu(pos^\epsilon(\{o^u_2\})) &=& 1
\end{eqnarray}
which shows that the inequalities of theorem 1 (see (8)) 
are very strong in this
quantum case. 

\subsection{The classical situation ($\epsilon = 0$)}
The classical situation is the situation without 
fluctuations. If $\epsilon$ = 0,
then $d$ can take any value in the interval $[-1,+1]$, and we 
have:
\begin{eqnarray}
eig^\epsilon(\{o^u_1\}) &=& \{ p_v \ \vert \  d < v \cdot u \} \\
eig^\epsilon(\{o^u_2\}) &=& \{p_v \
\vert \  v \cdot u < d \} \\
pos^\epsilon(\{o^u_1\}) &=& \{ p_v \ \vert \  d \le v \cdot u \} \\
pos^\epsilon(\{o^u_2\}) &=& \{p_v \
\vert \  v \cdot u \le d \}
\end{eqnarray}
We again consider the situation of an at random 
distribution of the point particle
on the sphere described by the probability distribution $\mu$. We 
then have in this case:
\begin{eqnarray}
P^\epsilon(\{o^u_1\}, \mu) &=& {1\over 2}(1-d) \\
P^\epsilon(\{o^u_2\}, \mu) &=& {1\over
2}(1+d) 
\end{eqnarray}
On the other hand we have:
\begin{eqnarray}
\mu(eig^\epsilon(\{o^u_1\})) &=& {1\over 2}(1-d) = 
\mu(pos^\epsilon(\{o^u_1\})) \\
\mu(eig^\epsilon(\{o^u_2\})) &=& {1\over 2}(1+d) = 
\mu(pos^\epsilon(\{o^u_2\}))
\end{eqnarray}
which shows that the inequalities of theorem 
\ref{theorem1} (see (\ref{equation1})) have become
equalities in this case.

\subsection{The general situation}
To give a clear picture of the general situation, we 
introduce additional concepts.
First we remark that the regions of eigenstates 
$eig^\epsilon(\{o^u_1\})$ and
$eig^\epsilon(\{o^u_2\})$, and the regions of possibility states 
$pos^\epsilon(\{o^u_1\})$ and
$pos^\epsilon(\{o^u_2\})$, are determined by the points of 
spherical sectors of $surf$ centered around
the points $u$ and $-u$ (see Fig. 4 and Fig. 5).

\vskip 0.7 cm
\hskip 3.2 cm \includegraphics{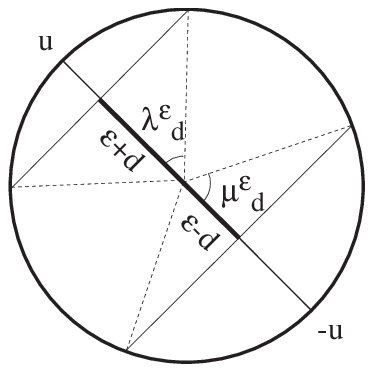}

\begin{quotation}
\noindent \baselineskip= 9 pt \smallroman Fig. 5 :  We have represented the 
different angles
$\scriptstyle \lambda^\epsilon_d$ and $\scriptstyle 
\mu^\epsilon_d$ characterizing the
spherical sectors of the eigenstate sets and possibility state 
sets.
\end{quotation}
We denote a closed spherical sector centered
around the point $u \in surf$ with angle $\theta$ by 
$sec(u,\theta)$, and a open spherical
sector with the same angle by $sec^o(u,\theta)$. We call 
$\lambda^\epsilon_d$ the angle of
the spherical sectors corresponding to $eig^\epsilon(\{o^u_1\})$ 
for all $u$, hence for $0 \not=
\epsilon$ we have $eig^\epsilon(\{o^u_1\}) = \{p_v\ \vert\ v \in 
sec(u,\lambda^\epsilon_d) \}$, and
$eig^\epsilon(\{o^u_1\}) = \{p_v\ \vert\ v \in sec^o(u, 
\lambda^0_d) \}$ for $\epsilon = 0$ (see Fig. 5). We can verify easily that $eig^\epsilon(\{o^u_2\})$ is 
determined by a spherical sector centered
around the point $-u$. We call $\mu^\epsilon_d$ the angle of this 
spherical sector, hence for
$0 \not= \epsilon$ we have $eig^\epsilon(\{o^u_2\}) = \{p_v\ 
\vert\ v \in sec(-u, \mu^\epsilon_d)\}$
and $eig^\epsilon(\{o^u_2\}) = \{p_v\ \vert\ v \in sec^o(-u, 
\mu^\epsilon_d) \}$ for $\epsilon = 0$.
For $0 \not= \epsilon$ we have $pos^\epsilon(\{o^u_1\}) = \{p_v\ 
\vert\ v \in
sec^o(u,\pi-\mu^\epsilon_d) \}$, and $pos^\epsilon(\{o^u_1\}) = 
\{p_v\ \vert\ v \in sec(u,
\pi-\mu^0_d) \}$ for $\epsilon = 0$. For $0 \not= \epsilon$ we
have $pos(\{o^u_2\}) = \{p_v\ \vert\ v \in sec^o(-u, \pi-
\lambda^\epsilon_d)\}$ and
$pos^\epsilon(\{o^u_2\}) = \{p_v\ \vert\ v \in sec(-u, \pi-
\lambda^\epsilon_d) \}$ for $\epsilon = 0$.
We have: 
\begin{eqnarray}
cos\lambda^\epsilon_d &=& \epsilon +d \\
cos\mu^\epsilon_d &=& \epsilon - d \\
\lambda^\epsilon_{-d} &=& \mu^\epsilon_d
\end{eqnarray}

\section{The classical situation}

We want to formulate now the classical situation in this 
general scheme. We shall
see that in the characterization of a classical probability model, 
the inequalities of
theorem 1 play an important role.

\begin{definition} \label{definition1} Suppose that we are in the 
situation $\mu$ of lack of
knowledge on the states. We say that an experiment $e$ is a 
'classical experiment' iff
$\mu(eig(A^e)) = \mu(pos(A^e))$ for all subsets $A^e \subset O_e$. 
With other words, a
classical experiment is an experiment where all states, except a 
collection of measure
zero, give rise to predetermined outcomes for this experiment.
\end{definition}
Classical experiments are experiments with predetermined 
outcomes (except for
a set of states of measure zero that as a consequence do not 
contribute to the
statistics). For these classical experiments we can show that 
probability always originates in
a lack of knowledge on the states.

\begin{theorem} If we are in the situation $\mu$ 
of lack of knowledge on the
states, and $e$ is a classical experiment, then for
each $A^e \in {\cal B}(O_e)$ we have :
\begin{equation}
\mu(eig(A^e)) = P(A^e, \mu) = \mu(pos(A^e))
\end{equation}
\end{theorem}

\noindent {\sl Proof:} A direct consequence of definition \ref{definition1} and 
theorem \ref{theorem1}.

\smallskip
\noindent We shall show now that for classical experiments Bayes 
formula for the conditional
probability is valid. To be able to analyze the validity of
Bayes formula in the scheme that we have presented here, we must 
give an operational
definition for the concept of conditional probability. Here we are 
confronted with a
conceptual problem, since in most textbooks, the conditional 
probability is defined by means
of Bayes formula. Since the conditional probability is a
primary physical quantity that is measured in the laboratory, we 
should define it
operationally and without the use of Bayes formula.

\section{The concept of conditional probability}

We want to make clear that there is a distinction 
between the occurrence
of an outcome when an experiment is performed, and the 
conditioning on an
outcome corresponding to an experiment.

\begin{definition} Given a situation $\mu$ of lack 
of knowledge on the
states of an entity $S$, described by a probability measure on 
this set of states
$\Sigma$. We condition the entity $S$ on a subset $A_f \subset 
O_f$ for an experiment $f$,
if we consider during the performance of the experiment $e$ only 
those trials, where
the situation of the entity before the experiment $e$ is such that 
we can predict the
outcome for the experiment $f$ to occur with certainty in $A_f$, 
if we would decide to
perform the experiment $f$.
\end{definition}
From this definition follows that
conditioning is equivalent to a change of the situation $\mu$ 
before the
experiment in such a way that the experiment $f$ would give with 
certainty an outcome in
$A_f$, if it would be executed. The new situation of lack of 
knowledge is described by the
probability measure that we shall denote by $\mu_{A_f} : {\cal 
B}(\Sigma) \to [0,1]$. It
is defined for an arbitrary subset $K \subset \Sigma$ as follows :

\begin{equation}
\mu_{A_f}(K) = \mu(K \cap eig(A_f)) / \mu(eig(A_f)) \label{equationconditional}
\end{equation}
Now that we have introduced this concept of 
`conditioning' on an experiment, we can
introduce the general concept of conditional probability.

\begin{definition} Given a situation 
$\mu$ of lack of
knowledge on the states of an entity, described by the probability 
measure $\mu$, and given
two experiments $e$ and $f$, then we want to consider the 
conditional probability $P(A_e,
A_f, \mu)$. This is the probability that the experiment $e$ makes 
occur an outcome in the set
$A_e$, when the situation is conditioned on the set $A_f$ for the 
experiment $f$. The
conditional probability is a map $P : {\cal B}(O_e) \times {\cal 
B}(O_f) \times {\cal
M}(\Sigma) \to [0,1]$.
\end{definition}

\begin{theorem} Given a situation $\mu$ of lack of 
knowledge on the states,
and two experiments $e$ and $f$. If the experiment $e$ is a 
classical experiment, then the
conditional probability $P(A_e, A_f, \mu)$ satisfies Bayes 
formula. More specifically we have
$P(A_e, A_f, \mu) = \mu(eig(A_e) \cap eig(A_f)) / \mu(eig(A_f))$.
\end{theorem}
{\sl Proof:} We have (see \ref{equationconditional}) that $P(A_e, A_f, \mu) = 
P(A_e, \mu_{A_f})$.
If $e$ is a classical experiment it follows from theorem 2 that 
$P(A_e, \mu_{A_f}) =
\mu_{A_f}(eig(A_e)) = \mu(eig(A_e) \cap eig(A_f)) / 
\mu(eig(A_f))$. This shows that Bayes
formula is valid.

\bigskip
\noindent  From this theorem can intuitively be seen that Bayes 
formula for the conditional
probability shall not be valid for the case of two experiments 
that are both non-classical. We
shall now analyze all these situations in the $\epsilon$-model and 
show how the conditional
probability in the $\epsilon$-model evolves continuously from the 
quantum transition
probability, for the case of $\epsilon = 1$, to a classical 
Kolmogorovian probability
satisfying Bayes formula, for the case $\epsilon = 0$. We shall 
also show that for values of
$\epsilon$ strictly between $1$ and $0$, the conditional 
probability is neither quantum nor
classical.

\section{The conditional probability and the $\epsilon$-
model}

Given a situation $\mu$ of lack of knowledge about the 
state of the point
particle, described by a uniform probability measure on the 
sphere. This corresponds to the
situation where the particle $P$ is distributed at random on the 
sphere. For a fixed
$\epsilon$, and two parameters $d$ and $c$ both in the interval 
$[-1+\epsilon, 1-\epsilon]$,
there are also given the two experiments $e^\epsilon_{u,d}$ and 
$e^\epsilon_{w,c}$.

\vskip 0.7 cm
\hskip 3.2 cm \includegraphics{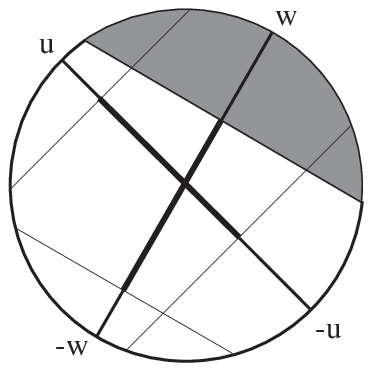}

\begin{quotation}
\noindent \baselineskip= 9 pt \smallroman Fig. 6 : An illustration of the 
situation corresponding
to the conditional probability for the $\scriptstyle \epsilon$-
model. The lack of
knowledge about the state of the particle is described by a 
uniform probability
distribution $\scriptstyle \mu$ on the sphere. For a fixed 
$\scriptstyle \epsilon$ and two
parameters $\scriptstyle d$ and $\scriptstyle c$ in the the 
interval $\scriptstyle
[-1+\epsilon, 1-\epsilon]$, we consider two experiments 
$\scriptstyle e^\epsilon_{u,d}$
and $\scriptstyle e^\epsilon_{w,c}$. We want to calculate the 
conditional probability
$\scriptstyle P(u, w, \mu)$ that the experiment $\scriptstyle 
e^\epsilon_{u,d}$ gives the
outcome $\scriptstyle o^u_1$, when the entity is conditioned on 
the outcome $\scriptstyle
o^w_1$ for the experiment $\scriptstyle e^\epsilon_{w,c}$. This 
means that before the start
of the experiment $\scriptstyle e^\epsilon_{u,d}$, the situation 
is such that if we would
perform the experiment $\scriptstyle e^\epsilon_{w,c}$, the 
outcome $\scriptstyle
o^w_1$ would come out with certainty. This conditioning is 
expressed by a new
probability measure on the sphere, which is zero outside the grey 
area, and equal to
the old one, except for a renormalization factor, in the grey 
area.
\end{quotation}
In general we consider the conditional probability for 
arbitrary elements of the set
of measurable subsets of the outcome sets of the two experiments. 
Since in the
$\epsilon$-model we only have experiments $e^\epsilon_{u,d}$, with 
two outcomes $o^u_1$ and
$o^u_2$, we want to alleviate somewhat the notation. Therefore we 
shall denote the
conditional probability that the experiment $e^\epsilon_{u,d}$ 
gives the outcome $o^u_1$
(respectively $o^u_2$), when the entity is conditioned for the 
outcome $o^w_1$ of the
experiment $e^\epsilon_{w,c}$ by $P(u, w, \mu)$ (respectively $P(-
u, w, \mu)$). In a similar
way we denote the conditional  probability that the experiment 
$e^\epsilon_{u,d}$ gives the outcome $o^u_1$
(respectively $o^u_2$), when the entity is conditioned for the 
outcome $o^w_2$ of the
experiment $e^\epsilon_{w,c}$ by $P(u, -w, \mu)$ (respectively 
$P(-u, -w, \mu)$) (see Fig. 6). We repeat again: the conditional probability $P(u, w, \mu)$ is 
the probability that
the experiment $e^\epsilon_{u,d}$ gives the outcome $o^u_1$, if 
the entity is conditioned on
the outcome $o^w_1$ for the experiment $e^\epsilon_{w,c}$. This 
means that the lack of
knowledge on the states is such that if we would decide to perform 
the experiment
$e^\epsilon_{w,c}$, the outcome $o^w_1$ would come out with 
certainty. With other words (see
Fig. 6), the state of the entity is such that the particle is 
distributed uniformly inside
the spherical sector $eig(\{o^w_1\})$, the grey area on Figure 6. 
It is easy to
formulate in a similar way the four other conditional 
probabilities $P(-u, w, \mu), P(u, -w,
\mu)$ and $P(-u, -w, \mu)$. The explicit calculation of these 
conditional probabilities is a
long exercise of classical calculus, and therefore we refer to 
(Aerts D. and Aerts S. 1994 b)
for a detailed exposition of this calculation. Here we only give 
the result. Let us call
$\alpha$ the angle between the two vectors $u$ and $w$, then we 
have: 

\begin{equation}
P(u, w, \mu) = p_1
\cdot H(\epsilon - cos{\alpha\over 2}) + H(\epsilon - 
sin{\alpha\over 2}) \cdot p_2 \cdot
H(cos{\alpha\over 2} - \epsilon) + p_3 \cdot H (sin{\alpha\over 2} 
- \epsilon) \label{equationp}
\end{equation}
where $H(x)$ is the function of Heaviside, and : 
\begin{equation}
p_1 = 
{cos\alpha (1+\epsilon)\over
{4\epsilon}} + {1\over 2}
\end{equation}
\begin{equation}
p_2 = p_1 + {1\over 2} + 
{\omega (u,w)\over
{4\pi(1-\epsilon)}} + {cos\alpha+1\over {4\pi\epsilon(1-
\epsilon)}} \cdot \sigma (u,w) 
\end{equation}
\begin{equation}
p_3 = p_1 + {\omega(u,w)-\omega(-u,w)\over 4\pi(1-
\epsilon)} + {(cos\alpha-1)
\cdot \sigma(-u,w) + (cos\alpha +1) \cdot \sigma(u,w)\over 
4\pi\epsilon(1-\epsilon)}
\end{equation}
where :
\begin{equation}
\omega(u,w) = 4\epsilon \cdot Arc cos\sqrt{{1 - ({\epsilon\over 
cos(\alpha/2)})^2}\over
{1-\epsilon^2}} - 4 Arcsin{sin(\alpha/2)\over {\sqrt{(1-
\epsilon^2}}}
\end{equation}
and :
\begin{equation}
\sigma(u,w) = \epsilon \cdot tg(\alpha/2) \cdot \sqrt{1-
({\epsilon\over cos(\alpha/2)})^2} -
(1-\epsilon^2) \cdot Arccos({\epsilon \cdot tg(\alpha/2)\over 
\sqrt{1-\epsilon^2}})
\end{equation}
Let us 
first
consider the two extremal cases, the classical case, where 
$\epsilon =0$, and the quantum
case, where $\epsilon =1$. 

\bigskip
\noindent {\it (1) the classical case ($\epsilon = 0$)}

\medskip
\noindent In this case we find :
\begin{equation}
P(u, w, \mu) = 1 - {\alpha\over \pi}
\end{equation}
which is a linear function in the angle.

\bigskip
\noindent {\it (2) the quantum case ($\epsilon = 1$)}

\medskip
\noindent For this case we only have to take into account the 
contribution $p_1$ of
(\ref{equationp}), and hence we find :
\begin{equation}
P(u, w, \mu) = cos^2(\alpha/2)
\end{equation}
which is the well known quantum transition probability between the 
states $p_u$ and $p_w$.

\smallskip
\noindent
The conditional probability $P(u, w, \mu)$ 
evolves continuously from
the quantum transition probability between the states $p_u$ and 
$p_w$ to a linear
function of the angle between the two vectors $u$ and $w$. 

\section{An intermediate situation of the $\epsilon$-
model that is neither
classical nor quantum}

A complete probabilistic analysis of the intermediate 
situation of the
$\epsilon$-model is presented in (Aerts D. and Aerts S. 1994 b). 
Here we only show that for a
specific value of $\epsilon = {\sqrt2\over 2}$ the conditional 
probabilities of the
$\epsilon$-model cannot be fitted neither into a quantum 
probability model nor into a
classical probability model. 

\begin{theorem} For $\epsilon =
{\sqrt2\over 2}$, the conditional probabilities of the $\epsilon$-
model cannot be fitted
neither into a classical nor into a quantum probability model.
\end{theorem}
{\sl Proof :} We shall give a proof ex absurdum, and 
suppose that there does
exist a Kolmogorovian model satisfying Bayes formula for the 
conditional probability for
this value of $\epsilon$. We consider three experiments
$e^\epsilon_{u,d}$, $e^\epsilon_{v,d}$ and $e^\epsilon_{w,d}$, 
such that $d = 0$ and $u$, $v$
and $w$ are in the same plane, with an angle of $2\pi\over 3$ 
between $u$ and $v$, between $v$
and $w$, and between $w$ and $u$. If we use the general expression
for the conditional probability (\ref{equationp}) we find $P(v,w,\mu) = 0.78$, 
$P(u,w,\mu) = 0.22$ and
$P(-u, v, \mu) = 0.22$. As we have defined, $P(v,w,\mu)$ is the 
probability that the
experiment $e^\epsilon_{v,d}$ gives the outcome $o^v_1$, if we 
have conditioned the entity in
such a way that if we would perform the experiment 
$e^\epsilon_{w,d}$, it would give with
certainty an outcome $o^w_1$. To be able to express more clearly 
the
hypothesis of the existence of a Kolmogorovian probability model, 
we write these
conditional probabilities in a more standard notation. Hence 
$P(v,w,\mu) = P(e^\epsilon_{v,d}
= o^v_1 \ \vert\ e^\epsilon_{w,d} = o^w_1)$, $P(u,w,\mu) = 
P(e^\epsilon_{u,d} = o^v_1 \ \vert\
e^\epsilon_{w,d} = o^w_1)$, and $P(-u, v, \mu) = 
P(e^\epsilon_{u,d} = o^v_2 \ \vert\
e^\epsilon_{v,d} = o^v_1)$ under the preparation $\mu$. If there 
does exist a
Kolmogorovian probability model, satisfying the Bayes formula, 
there exists a
set $X$ (playing the role of the sample space), and a $\sigma$-
algebra ${\cal B}(X)$ (playing
the role of the set of events), and a probability measure $\nu : 
{\cal B}(X) \to [0,1]$, such
that the conditional probabilities $P(e^\epsilon_{v,d} = o^v_1 \ 
\vert\ e^\epsilon_{w,d} =
o^w_1)$, $P(e^\epsilon_{u,d} = o^v_1 \ \vert\ e^\epsilon_{w,d} = 
o^w_1)$ and
$P(e^\epsilon_{u,d} = o^v_2 \ \vert\ e^\epsilon_{v,d} = o^v_1)$ 
can be written under the
appropriate form. This means that there exists elements $U, V, W 
\in {\cal B}(X)$ such that :
\begin{eqnarray}
P(e^\epsilon_{v,d} = o^v_1 \ \vert\ e^\epsilon_{w,d} = o^w_1) &=& 
{\nu(V \cap W)\over \nu(W)} \\
P(e^\epsilon_{u,d} = o^v_1 \ \vert\ e^\epsilon_{w,d} 
= o^w_1) &=& {\nu(U \cap
W)\over \nu(W)} \label{equation30} \\
P(e^\epsilon_{u,d} = o^v_2 \ \vert\ e^\epsilon_{v,d} = o^v_1) &=& 
{\nu(U^C \cap V)\over
\nu(V)} \label{equation31}
\end{eqnarray}
  We also have $\nu(W) = \mu(o^w_1,
\mu) = {1\over 2}$ and $\nu(V) = \mu(o^v_1, \mu) = {1\over 2}$. 
Using the
fact that $\nu$ is a probability measure, and (\ref{equation30}) and (\ref{equation31}), and 
substituting the values of
the conditional probabilities, we get: 
\begin{eqnarray}
\nu(V \cap W) &=& \nu(U
\cap V \cap W) + \nu (U^C \cap V \cap W) \\
&=& {1\over 2} \cdot 
P(v,w,\mu) = 0.39 \label{equation32} \\
\nu(U \cap W) &=& \nu(U \cap V \cap W) + \nu (U \cap V^C \cap W) \\
&=& {1\over 2}
\cdot P(u,w,\mu) = 0.11 \label{equation33} \\
\nu(U^C \cap V)
&=& \nu(U^C \cap V \cap W) + \nu(U^C \cap V \cap W^C) \\
&=& {1\over 2} 
\cdot P(-u,v,\mu) = 0.11 \label{equation34}
\end{eqnarray}
If we subtract (\ref{equation33}) from (\ref{equation32}) we find :
\begin{equation}
\nu(U^C \cap V \cap W) = 0.28 + \nu(U \cap V^C \cap W) 
\end{equation}
which implies that
\begin{equation}
0.28 \le \nu(U^C \cap V \cap W) \label{equation42}
\end{equation}
On the other hand from (\ref{equation34}) follows 
\begin{equation}
\nu(U^C \cap V \cap W) \le 0.11 \label{equation43}
\end{equation} 
The inequalities (\ref{equation42}) and (\ref{equation43}) deliver us the contradiction that 
we were looking for. The
conclusion is that for these values of the conditional 
probabilities there does not exists a
Kolmogorovian probability model satisfying the Bayes formula. 

Also for the proof of the non-existence of a Hilbertian 
model we proceed ex
absurdum. If there exists a two dimensional complex Hilbert space 
model, where the
conditional probabilities are described by the transition 
probabilities, we can find three
orthonormal basis $\{\phi_1, \phi_2\}$, $\{\psi_1, \psi_2\}$, 
$\{\chi_1, \chi_2\}$ such
that: 
\begin{eqnarray}
\gamma^2 &=& \vert \langle \phi_1, \psi_1\rangle \vert^2 = 
\vert \langle \psi_1,
\chi_1\rangle \vert^2 = \vert \langle \chi_1, \phi_2\rangle 
\vert^2 \\
&=& \vert \langle \phi_2,
\psi_2\rangle \vert^2 = \vert \langle \psi_2, \chi_2\rangle 
\vert^2 = \vert \langle \chi_2,
\phi_1\rangle \vert^2 = 0.78 \\
\delta^2 &=& \vert 
\langle \phi_1, \psi_2\rangle
\vert^2 = \vert \langle \psi_1, \chi_2\rangle \vert^2 = \vert 
\langle \chi_1, \phi_1\rangle
\vert^2 \\
&=& \vert \langle \phi_2, \psi_1\rangle \vert^2 = \vert 
\langle \psi_2, \chi_1\rangle
\vert^2 = \vert \langle \chi_2, \phi_2\rangle \vert^2 = 0.22
\end{eqnarray}
This means that
there exists five angles $\theta_1, \theta_2, \theta_3, \theta_4$ 
and $\theta_5$ such that:

\begin{eqnarray}
\langle \chi_1, \psi_2 \rangle &=& \delta \cdot e^{i\theta_1} \\
 \langle \chi_1, \phi_1 \rangle
&=& \delta \cdot e^{i\theta_2} \\
\langle \phi_1, \psi_2 \rangle &=& 
\delta \cdot
e^{i\theta_3} \\
\langle \chi_1, \phi_2 \rangle &=& \gamma \cdot 
e^{i\theta_4} \\
\langle \phi_2,
\psi_2 \rangle &=& \gamma \cdot e^{i\theta_5}
\end{eqnarray}
If $\{\phi_1, \phi_2\}$ is an orthonormal basis, we have $\langle 
\chi_1, \psi_2 \rangle =
\langle \chi_1, \phi_1 \rangle \langle \phi_1, \psi_2 \rangle + 
\langle \chi_1, \phi_2
\rangle \langle \phi_2, \psi_2 \rangle $, and hence
\begin{equation}
\delta \cdot e^{i\theta_1} = \delta
\cdot e^{i\theta_2} \cdot \delta \cdot e^{i\theta_3} + \gamma 
\cdot e^{i\theta_4} \cdot \gamma
\cdot e^{i\theta_5}
\end{equation}
and also the complex conjugate
\begin{equation}
\delta \cdot e^{-i\theta_1} = \delta \cdot e^{-i\theta_2} \cdot 
\delta \cdot e^{-i\theta_3}
+ \gamma \cdot e^{-i\theta_4} \cdot \gamma \cdot e^{-i\theta_5} 
\end{equation}
If we multiply these last equations term by term we find :
\begin{equation}
\delta^2 = \delta^4 + \gamma^4 + \delta^2\gamma^2 \cdot 
e^{i(\theta_2 + \theta_3 -
\theta_4 - \theta_5))} + \delta^2\gamma^2 \cdot e^{-
i(\theta_2+\theta_3-\theta_4-\theta_5))}
\end{equation}
This we can write as
\begin{equation}
\delta^2 = \delta^4 + \gamma^4 + 2\delta^2\gamma^2 \cdot 
cos(\theta_2+\theta_3 - \theta_4
- \theta_5)
\end{equation}
But then we must have
\begin{equation}
cos(\theta_2+\theta_3-\theta_4-\theta_5) = {\delta^2-\delta^4-
\gamma^4\over
{2\delta^2\gamma^2}}
\end{equation}
If we fill in the values $\delta^2 = 0.22$ and $\gamma^2 = 0.78$ 
we find :
\begin{equation}
cos(\theta_2+\theta_3-\theta_4-\theta_5) = -1.27
\end{equation}
From this contradiction we can conclude that there does not exist 
a two dimensional
Hilbert space model, such that the conditional probabilities can 
be described by transition
probabilities in this Hilbert space.
\smallskip
\noindent
This theorem shows that we really have identified a new region of 
probabilistic structure in
this intermediate domain. In (Aerts D. and Aerts S. 1994 b) we 
show that for any value
of $\epsilon$ different from $0$, the probability structure of the 
$\epsilon$-model is non
Kolmogorovian (not satisfying Bayes formula for the conditional 
probabilities). We
also show that there is a domain of $\epsilon \not= 1$ where a 
Hilbert
space model can be found, but another domain where this is not the 
case.

We come now to the last section of this paper, where we like to 
give a sketch of how these
non-Kolmogorovian probabilities appear in other regions of nature.

\section{Non-Kolmogorovian probabilities in other regions 
of nature}

As follows from the foregoing analysis, non-classical 
experiments, giving rise
to a non-classical structure, are characterized by the presence of 
non-predetermined outcomes.
This makes it rather easy to recognize the non-classical aspects 
of experiments in other
regions of reality. Let us consider the situation of a decision 
process developing in the
mind of a human being, and we refer to (Aerts D. and Aerts. S. 
1994 a, b) for a more detailed
description. Hence our entity is a person, its states being the 
possible 'states' of this
person. Experiments are questions that can be asked to the person, 
and on which she or he has
to respond with 'yes' or 'no'. The typical situation of an opinion 
poll can be thought of as a
concrete example. Let us consider three different
questions : 

\begin{enumerate}

\item $u$ : Are you in favor of 
the use of nuclear energy?

\item $v$ : Do you think it would be a good 
idea to legalize soft-drugs?
\smallskip

\item $w$ : Do you think capitalism is better than 
social-democracy?
\end{enumerate}

We have chosen such type of questions that many persons 
shall not have
predetermined opinions about them. Since the person has to respond 
with 'yes' or 'no', she or
he, without opinion before the questioning, shall 'form' her or 
his opinion
during this process of questioning. We can use the $\epsilon$-
model to represent this
situation.

\vskip 0.7 cm
\hskip 0.7 cm \includegraphics{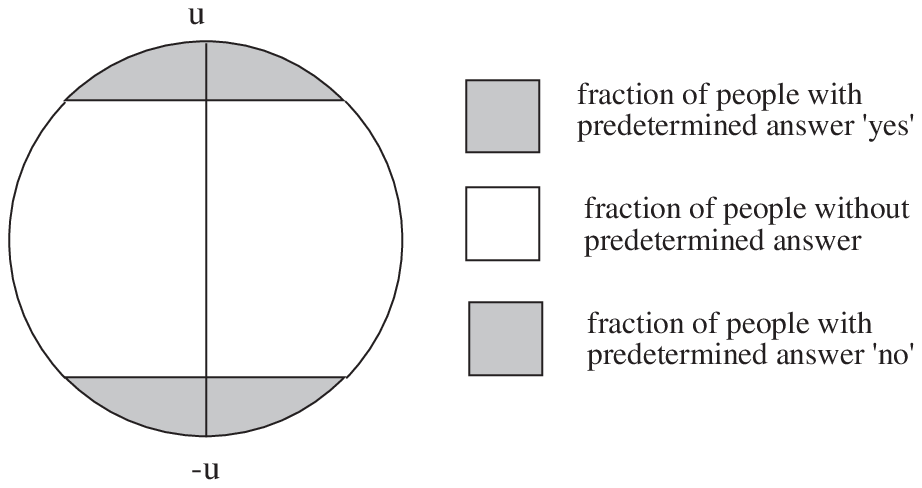}

\begin{quotation}
\noindent \baselineskip= 9 pt \smallroman Fig. 7 : A representation of the 
question $\scriptstyle
u$ by means of the $\scriptstyle \epsilon$-model. We have 
indicated the three regions
corresponding to predetermined answer 'yes', without predetermined 
answer, and predetermined
answer 'no'.
\end{quotation}
To simplify the situation, but without touching the 
essence, we make the
following assumptions about the probabilities that are involved. 
We suppose that in all cases
50\% of the persons have answered the question $u$ with 'yes', but 
only 15\% of the persons
had a predetermined opinion. This means that 70\% of
the persons formed their answer during the process of questioning. 
For simplicity we make the
same assumptions for $v$ and $w$. We can represent this situation 
in the $\epsilon$-model as
shown in Figure 7. We also make some assumptions of the way in 
which the different
opinions related to the three questions influence each other. We 
can represent an example of a
possible interaction by means of the $\epsilon$-model (Figure 9). 
One can see how a person
can be a strong proponent for the use of nuclear energy, while 
having no predetermined
opinion about the legalization of soft drugs (area 1 in Figure 9). 
Area (4) corresponds to
a sample of persons that have predetermined opinion in favor of 
legalization of soft drugs
and in favor of capitalism. For area (10) we have persons that 
have
predetermined opinion against the legalization of soft drugs and 
against capitalism. All the
13 areas of Figure 9 can be described in such a simple way.

\vskip 0.7 cm
\hskip 3 cm \includegraphics{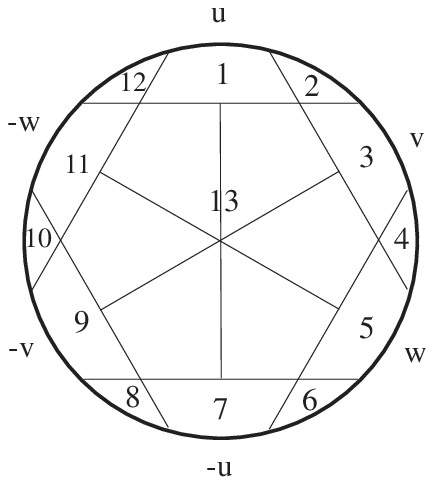}

\begin{quotation}
\noindent \baselineskip= 9 pt \smallroman Fig. 8 : A representation of the 
three questions $\scriptstyle
u$, $\scriptstyle v$, and $\scriptstyle w$  by means of the 
$\scriptstyle \epsilon$-model. We
have numbered the 13 different regions. For example: (1) 
corresponds to a sample of persons
that have predetermined opinion in favor of nuclear energy, but 
have no predetermined
opinion for both other questions, (4) corresponds to a sample of 
persons that have
predetermined opinion in favor of legalization of soft drugs and 
in favor of capitalism,
(10) corresponds to a sample of persons that have predetermined 
opinion against the
legalization of soft drugs and against capitalism, (13) 
corresponds to the sample of
persons that have no predetermined opinion about non of the three 
questions, etc... 
\end{quotation}
Deliberately we have chosen the different fractions of 
people in such a way that
the conditional probabilities fit into the $\epsilon$-model for a 
value of $\epsilon =
{\sqrt2\over 2}$. This means that we can apply theorem 4, and 
conclude that the collection of
conditional probabilities corresponding to these questions $u$, 
$v$ and
$w$ can neither be fitted into a Kolmogorovian probability model 
nor into a quantum
probability model. We are developing now in Brussels a statistics 
for such new
situations, that we have called 'interactive statistics'. By means 
of this statistics it
should be possible to make models for situations where part of the 
properties to be tested
are created during the process of testing. 

\section{Conclusion}

The further development of an intermediate (between classical and 
quantum) probability theory
and an interactive statistics could be very fruitful as well for 
physics as for other
sciences.

\par We work now at the construction of a general theory, where 
also the intermediate
structures can be incoorporated (Aerts 1986, Aerts 1987, Aerts, 
Durt and
Van Bogaert 1993, Aerts 1994, Aerts and Durt 1994a, 1994b, Coecke 
1994a, 1994b, 1994c). This
theory can probably be used to describe the region of reality 
between microscopic and
macroscopic, often referred to as the mesoscopic region. Actually, 
physicists use a very
complicated heuristic mixture of quantum and classical theories to 
construct models for
mesoscopic entities. There is however no consistent theory, and a 
general intermediate theory
could perhaps fill this gap. We try to find examples of simple 
physical phenomena in the
mesoscopic region that could eventually be modeled by an 
$\epsilon$-like model (Aerts and Durt
1994 b). If we succeed in building this intermediate theory, we 
would not only have a new
theory for the mesoscopic region, but the existence of such a 
theory would also shed light on
old problems of quantum mechanics (the quantum classical relation, 
the classical limit, the
measurement problem, etc...). In  

Till now we have only been developing the
kinematics of this intermediate theory (Aerts 1994 and Aerts and 
Durt 1994 a), but, once the
kinematics is fully developed, the way to construct a dynamics for 
the intermediate
region is straightforward. We can study the imprimitivity system 
related to
the Galilei group and look for representations of this Galilei 
group in the
group of automorphisms of the kinematical structure of the 
intermediate theory. If we can
derive an evolution equation in this way, it should continuously 
transform with varying
$\epsilon$ from the Schr\"odinger equation ($\epsilon = 1$), to 
the Hamilton
equations ($\epsilon =0$). 

\par As we have shown in the last section of this paper, the 
development of an interactive
statistics could be of great importance for the human sciences, 
where often
non-predetermined outcome situations appear. It could lead to a 
new methodology for these
sciences. Actually one is aware of the problem of the interaction 
between subject and
object, but it is generally thought that this problems cannot be 
taken into account in the
theory.

\par We also want to remark that the 'hidden measurement approach' 
defines a new quantization
procedure. Starting from a classical mechanical entity, and adding 
'lack of knowledge
(or fluctuations)' on the measurement process, out of the 
classical entity appears a
quantum-like entity. Another problem that we are investigating at 
this moment is an attempt
to describe quantum chaos by means of this new quantization 
procedure. It can be shown
that the sensitive dependence on the initial conditions that can 
be found in
the $\epsilon$-model for the classical situation in the set of 
unstable equilibrium
states disappears when the fluctuations on the measurement process 
increase. This could be
the explanation for the absence of quantum chaos.

\section{References}

\begin{description}

\item Accardi, L. (1982), {\it On the statistical meaning of 
the complex numbers in
quantum mechanics,} Nuovo Cimento, {\bf 34}, 161. 
 
\item Aerts, D. (1986), {\it A possible explanation for the 
probabilities of quantum
mechanics,} {\it J. Math. Phys.} {\bf 27}, 202.

 \item Aerts, D. (1987), {\it The origin of the non-
classical character of the
quantum probability model,} in {\it Information, Complexity, and 
Control in Quantum Physics}, 
A. Blanquiere, et al., eds., Springer-Verlag.

 \item Aerts, D. (1991), {\it A macroscopic classical 
laboratory situation with
only macroscopic classical entities giving rise to a quantum 
mechanical probability model,}
in {\it Quantum Probability and Related Topics, Vol. VI,} L. 
Accardi, ed., World Scientific,
Singapore.

 \item Aerts, D. and Van Bogaert, B. (1992), {\it A 
mechanical classical laboratory
situation with a quantum logic structure}, {\it Int. J. Theor. 
Phys.} {\bf 10}, 1893.

 \item Aerts, D.,  Durt, T. and  Van
Bogaert, B. (1992), {\it A physical example of quantum fuzzy sets, 
and the classical limit,}
in {\it Proceedings of the first International Conference on Fuzzy 
Sets and their
Applications, Tatra Montains Math. Publ.} {\bf 1}, 5.

\item Aerts, D., Durt, T. and Van Bogaert, B. (1993), 
{\it Quantum
probability, the classical limit and non-locality,} in {\it 
Symposium on the Foundations of
Modern Physics}, T. Hyvonen, ed., World Scientific, Singapore.

\item Aerts, D., Durt, T., Grib, A.A., Van Bogaert, B. and 
Zapatrin, R.R.
(1993), {\it Quantum structures in macroscopic reality,} {\it Int. 
J. Theor. Phys.} {\bf
32}, 3, 489.

\item Aerts, D. (1994), {\it Quantum structures, 
separated physical entities and
probability,} {\it Found. Phys.} {\bf 24}, 1227.

\item Aerts, D. and Durt, T. (1994 a), {\it Quantum, 
Classical and Intermediate, an
illustrative example,} {\it Found. Phys.} {\bf 24}, 1353.

\item Aerts, D and Durt, T. (1994 b), {\it Quantum, 
Classical and Intermediate; a
measurement model,}, in the proceedings of the {\it Symposium on 
the Foundations of Modern Physics,}
Helsinki, 1994, ed. K.V. Laurikainen, C.
Montonen, K. Sunnar Borg, Editions Frontieres, Gives sur Yvettes, France.

\item Aerts, D. and Aerts, S. (1994 a), {\it Applications of 
Quantum Statistics in
Psychological Studies of Decision Processes,} to be published in 
{\it Foundations of
Science,} {\bf 1}.

\item Aerts, D. and Aerts, S. (1994 b), {\it Interactive 
probability models: from
quantum to Kolmogorovian,} preprint, TENA, VUB, Pleinlaan 2, 1050 
Brussels, Belgium.

\item Birkhoff, G. and Von Neumann,(1936), J., Ann. Math. 
{\bf 37}, 823.

\item Coecke, B., (1994a), {\it A Hidden Measurement Model 
for Pure and Mixed
states of quantum mechanics in Euclidean space,} Int. J. Theor. 
Phys. (submitted to this
issue).

\item Coecke, B., (1994b), {\it Hidden measurement 
representation for quantum entities
described by finite dimensional complex Hilbert spaces,} preprint, 
TENA, VUB, Pleinlaan 2,
1050 Brussels, Belgium.

\item Coecke, B., (1994c), {\it A representation of pure and 
mixed states of quantum
physics in Euclidean space,} preprint, TENA, VUB, Pleinlaan 2, 
1050 Brussels, Belgium.

\item Emch, G. G. (1984), {\it Mathematical and conceptual 
foundations of 20th century
physics,} North-Holland, Amsterdam.

\item Foulis D., Randall C.,(1972), J. Math. Phys., 1667.

\item Gudder, S. P., (1988), {\it Quantum Probability,} 
Academic Press, Inc. Harcourt
Brave Jovanovich, Publishers. 

\item Jauch, J. M.,(1968), {\it  Foundations of Quantum 
Mechanics}, Addison-Wesley. 

\item Piron, C., (1976), {\it Foundations of Quantum 
Physics}, Benjamin, New York.

\item Pitovski, I. (1989) , {\it Quantum Probability - 
Quantum Logic,} Springer Verlag. 

\item Randall, C. and Foulis, D. (1979), {\it The 
operational approach to quantum
mechanics,} in {\it Physical theory as logico-operational 
structure,} ed. Hooker, C. A.,
Reidel.

\item Randall, C. and Foulis, D. (1983), {\it A mathematical 
language for
quantum physics}, in {\it Les Fondements de la M\'ecanique 
Quantique}, ed. C.
Gruber et al, A.V.C.P., case postale 101, 1015 Lausanne, Suisse.

\item Segal, I. E. (1947), Ann. Math. {\bf 48}, 930.

\end{description}

\end{document}